\documentclass[10pt,letterpaper]{article}
\usepackage{opex3,cite}
\usepackage{color}

\newcommand{\red}{\textcolor{black}}
\begin{document}
\title{Photoacoustic signal enhancements from gold nano-colloidal suspensions excited by a pair of time-delayed femtosecond pulses}
\author{Frances Camille P. Masim,$^1$ Wei-Hung Hsu,$^1$ Hao-Li Liu,$^2$
Tetsu Yonezawa,$^3$ Armandas Bal\v{c}ytis,$^{4,5}$ Saulius
Juodkazis,$^{4,6}$ and Koji Hatanaka$^{1,*}$}
\address{$^1$Research Center for Applied Sciences, Academia Sinica, 128 Sec. 2, Academia Rd., Nankang, Taipei 115, Taiwan (R. O. C.) \\
$^2$Department of Electrical Engineering, Chang Gung University,
259, Wenhua 1st Rd., Guishan Dist., Taoyuan 333, Taiwan (R. O. C.) \\
$^3$Division of Materials Science and Engineering, Faculty of
Engineering, Hokkaido University, Kita 13 Nishi 8, Kita-ku,
Sapporo, 0608628, Japan\\  $^4$Nanotechnology Facility, Center for
Micro-Photonics, Swinburne
University of Technology, John St., Hawthorn, VIC 3122, Australia\\  $^5$Center for Physical Sciences and Technology, Savanori\c{u} Ave. 231, LT-02300 Vilnius, Lithuania\\
$^6$Melbourne Centre for Nanofabrication, the Victorian Node of
the Australian National Fabrication Facility, 151 Wellington Rd.,
Clayton 3168 Vic, Australia }
\email{$^*$kojihtnk@gate.sinica.edu.tw}

\begin{abstract}
Photoacoustic signal enhancements were observed with a pair of
time-delayed femtosecond pulses upon excitation \red{of} gold
nanosphere colloidal suspension. A systematic experimental
investigation \red{of photoacoustic intensity} within the delay
time, $\Delta$t = 0 to 15~ns, was carried out. The results
revealed a significant enhancement factor of $\sim 2$ when the
pre-pulse energy is 20-30\% of the total energy. Pre-pulse and
main pulse energy ratios, $E_p^{(1)}$:$E_s^{(2)}$, were varied to
determine the optimal ratio that yields to maximum photoacoustic
signal enhancement. This enhancement was ascribed to the initial
stage of thermalization and \red{bubble} generation in the
nanosecond time scale. \red{P}re-pulse scattering intensity
measurements and \red{numerical finite-difference time-domain}
calculations were performed \red{to reveal dynamics and light
field enchancement, respectively}.
\end{abstract}

%%%%%%%%%%%%%%%%%%%%%%% References %%%%%%%%%%%%%%%%%%%%%%%%%
\bibliographystyle{osajnl}
%\bibliographystyle{spiebib}
%\bibliography{refs,new1,lpr,D:/papers/mybib/paper5a1}
%\bibliography{lpr,paper6a,proceedings}

%\maketitle
%\tableofcontents \newpage
%``''
\section{Introduction}

Recent studies on femtosecond (fs) laser-induced dielectric
breakdown have gained a widespread interest in the fabrication of
micro- and nano-functional devices~\cite{Du,Kumada,Ou} and
real-time high resolution imaging due to its high emission
frequency~\cite{ Danworaphong}. Fs-laser ablation offers
significant advantages over nanosecond laser ablation such as a
high precision with less of thermal damage and a high
reproducibility~\cite{Mildner, Penczak}. On this basis, fs-laser
pulses are suitable to improve high spatial resolution and
sensitivity in biomedical photoacoustic imaging and photothermal
therapy~\cite{Somekawa, Babushok}. Photoacoustic technique is one
of the most promising biophotonic diagnostic modalities that
incorporates non-ionizing radiation, non-invasive imaging, high
spatial resolution, and deep penetration depth~\cite{McLaughlan,
Dove, Jeon}. Photoacoustic signals are usually generated by one of
the four mechanisms: thermal expansion, vaporization, chemical
reaction induced by light, or optically induced dielectric
breakdown~\cite{Moon}. Thermal expansion that is accompanied by
cavitation generation results in efficient photoacoustic signal
generation~\cite{Wei,Huynh,Wilson}. \red{Formation mechanisms and
dynamics of nano-bubbles around nanoparticles under pulsed laser
irradiation are better understood following high resolution
optical and X-ray
imaging~\cite{Plech2004,Siems2011,Metwally2015,Boutopoulos2015} and
opened applications in cell laser optoporation and photothermal
therapy~\cite{Boulais2016,Lachaine2016}.}

The existing high-resolution optical imaging modalities such as
confocal microscopy, two-photon microscopy, and optical coherence
tomography limit its applications on deep penetration imaging due
to optical scattering. Compared with these imaging techniques,
photoacoustic imaging is known to surpass the optical diffusion
limit, providing a deeper penetration imaging with high spatial
resolution~\cite{Maslov,Wong,Urban}. It is considered to be a
potential imaging tool in neuroscience which allows penetration in
thick brain tissue. In vivo studies on noninvasive transdermal and
transcranial imaging of the structure and function of rat brains
by laser-induced photoacoustic tomography were reported~\cite{Rao,
Pang}. It allows accurate mapping of brain structures and
functional cerebral hemodynamic changes in blood vessels. This
neuroimaging modality is promising for significant applications in
neurophysiology, neuropathology and neurotherapy.

Plasmonic gold nanoparticles are attractive in photoacoustics
since they offer strong optical absorption when excited at the
surface plasmon resonance wavelength. Nanoparticle-facilitated
absorption of pulsed laser leads to a rapid and localized heating,
which results in photoacoustic signal generation produced through
thermo-elastic effect and cavitation generation~\cite{Muramatsu,
Lu, Zhang, Zhao, Hu}. The efficiency of optical absorption and
photothermal conversion can be tuned through nanoparticle
chemistry and geometry~\cite{Masim1}. It has been demonstrated
that enhanced photoacoustic intensity was observed by tuning the
nanoparticle shape~\cite{Masim2} and controlling the laser
parameters such as pulse energy and temporal chirp~\cite{Masim3}.

However, with the use of single fs-laser pulse, the pulse
parameters that influence the ablation process are circumscribed
to the pulse energy and pulse width. To achieve effective control,
double pulse, which contains two-polarized fs-pulses separated
from femtoseconds to nanoseconds, has been widely used in control
of light-matter interaction~\cite{Wang, Chen, Vogt}. It was found
that ablation can be precisely controlled by optimizing the number
of pulses, pulse separation, and pulse energy ratio~\cite{Pinon1,
Pinon2}. Reports on semiconductor materials revealed that the
ablation rate is higher for double-pulse compared to single-pulse
irradiation of the same total fluence~\cite{Guo,Harilal}. Double
pulse excitation leads to a better coupling of the laser beam with
plasma and target material, thus providing a more temporally
effective energy delivery to plasma and target material. This
results in significant signal enhancements in the intensity
emission lines up to two orders of magnitude larger than a
conventional single pulse excitation~\cite{Benedetti,De
Giacomo,Scaffidi}.

\red{Here, photoacoustic signal enhancements under double-pulsed
(horizontally- and vertically-polarized) excitation to Au
nanosphere colloidal suspensions were systematically
investigated.} Pre-pulse power dependence and different pre-pulse
to main pulse energy ratios with the same total fluence were
studied. The experimental results demonstrated that maximum
enhancement is obtained at the optimal separation time between
pulses and pulse energy ratios.

%__________________________________Fig. 1
\begin{figure}[tb]
\begin{center}
\includegraphics[width=14cm]{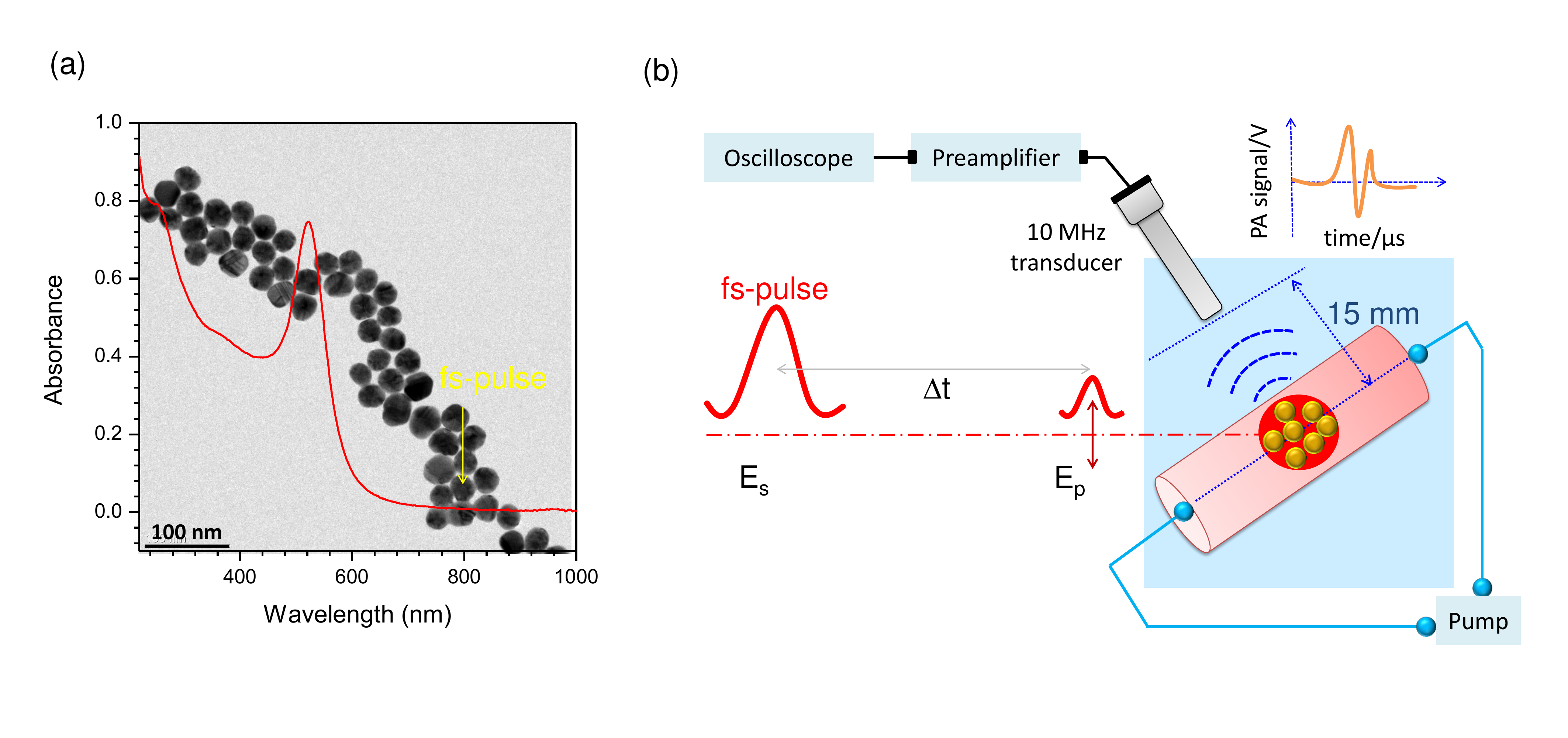}
\caption{(a) Absorption spectrum of Au nanosphere colloidal
suspension with a strong characteristic absorption band at
$\sim$520~nm. TEM image of mono-dispersed colloidal suspension of
Au nanosphere with a diameter of 20~nm is shown in the background.
(b) Schematic diagram of the experimental setup for
fs-double-polarized pulsed excitation to Au nanoparticle
suspension for photoacoustic detection. Fs-laser pulses: pulse
duration $t_0 = 40$~fs, central wavelength $\lambda = 800$~nm and
pulse energy $E = 0.1$~mJ at 1~kHz repetition rate were focused
inside the glass tube using $10\times$ numerical aperture $NA =
0.28$ objective lens. The pre-pulse was p-polarized while the main
pulse was s-polarized. Distance between the focal spot and
transducer was set at 15~mm in all experiments.} \label{f-a}
\end{center}
\end{figure}

\section{Samples and procedures}

\subsection{Synthesis of Au nanospheres}

\red{Colloidal suspensions of gold nanospheres for photoacoustic
generation were prepared via synthesis described
elsewhere~\cite{Bastus,Jana}}. Briefly, a
kinetically-controlled seeded growth synthesis of
citrate-stabilized gold nanospheres was used. In a 250~mL
three-necked round-bottomed flask, a solution of 2.2~mM sodium
citrate in Milli-Q water (150~mL) was heated at 115$^{\circ}$C for
15~min under vigorous stirring. A reflux condenser and an oil bath
were used to prevent the evaporation of the solvent. After it
reached the boiling point, 1~mL of HAuCl$_{4(aq)}$ (25~mM) was
added. The color of the solution changed from yellow to bluish
gray and then to light pink in 10~min. The resulting Au seed
particles $\sim$10~nm in diameter were coated with negatively
charged citrate ions and completely dispersed in water.
Immediately after the synthesis of the Au seed solution, the
temperature was cooled down to 90$^{\circ}$C and seeded growth of
Au nanospheres was carried out. Then, 1~mL of HAuCl$_{4(aq)}$
solution (25~mM) was injected on the reaction vessel. The reaction
was finished after 30~min and the process was repeated twice.
After that, the sample was diluted by extracting 55~mL of the
sample and adding 53~mL of Milli-Q water and 2~mL of 60~mM sodium
citrate. This solution was then used as seed solution, and the
process was repeated again. By changing the volume in each growth
step, it is possible to tune the seed particle concentration.
Mono-dispersed gold nanosphere colloidal suspensions with an
absorption peak at $\sim$520~nm corresponding to the diameter of
20~nm as shown in Fig.~\ref{f-a}(a). Atomic concentration of $\sim
1.4\times 10^{-4}$~mol/L, particle concentration of \red{$\sim
3.5\times 10^{14}$~NPs/L} and volume of $\sim 4\times 10^3$~nm$^3$
were used in the experiments. \red{Separation between particles
estimated as a cubic root of the volume-per-nanoparticle was $\sim
1.4~\mu$m. At this high-density, the formation and evolution of a
nano-bubbles is directly affected by pressure waves encountered
from surrounded nano-bubbles~\cite{Nakajima2016}.}

\subsection{Femtosecond double-pulse configuration}

Figure~\ref{f-a}(b) shows a schematic diagram of the experimental
setup of fs double-pulse experiments conducted using Ti:sapphire
amplified laser system with pulse duration of $t_0 = 40\pm 5$~fs,
central wavelength of $\lambda = 800$~nm, and pulse energy of $E_p
= 0.1$~mJ at 1~kHz repetition rate. Fs-laser pulses were directed
through two cube polarizing beam splitters in order to create a
pulse pair with s- and p-polarized beams. The first cube
polarizing beamsplitter split the incoming laser pulse into two
orthogonal optical paths while the second one was used to combine
the beams after introducing time delay $\Delta t$ between them in
one of the arms. The delay was controlled automatically by
mechanical stage with high precision. \red{The maximum delay time
range between the s- and p-polarized beams was 15~ns.} A half
waveplate which allows the rotation of the polarization vector was
used to control the intensity ratio between the two polarized
beams. \red{The minimum s-pol. to p-pol. beam intensity ratio is
50/50, therefore measurements on pulse energy ratios with the same
total fluence were restricted to 50/50, 60/40, 70/30, 80/20,
90/10, and 100/0.} After the second cube polarizing beam splitter,
the two polarized beams become collinear and were focused using
$10^\times$ numerical aperture $NA = 0.28$ objective lens.
\red{Two independent switches for s- and p-pol. beams were
installed to control excitation under single pulse and
double-pulse irradiation. For instance, in s-pol. beam
irradiation, the p-pol. beam was blocked and vice versa. In
contrary, under double-pulse excitation, both beams are present.}

\subsection{Photoacoustic detection and measurements}

\red{Femtosecond double-pulse with a total laser fluence of
$1.05\times 10^3$~J/cm$^2$ were tightly-focused onto a
5-mm-diameter glass capillary tube inside the water tank which is
used to circulate colloidal suspensions of gold nanoparticles. An
off-resonance (laser wavelength $\lambda = 800$~nm does not
coincide with the characteristic absorption band of Au nanosphere
at  $\lambda = 520$~nm) pulsed laser excitation of gold nanosphere
colloidal suspensions was performed, proving a higher thermal
stability of gold nanoparticles. For the photoacoustic detection
and measurements, a single element unfocused ultrasound transducer
(A312-N-SU) with a detection frequency of $10$~MHz was used.} The
distance between the ultrasound transducer and glass tube was kept
constant at 15~mm in the entire experiment. Photoacoustic signals
were detected and amplified using an ultrasound preamplifier
(5678, Olympus) and the acquired signals were recorded and
analyzed using digital oscilloscope (DSO-X 3034A, Agilent Tech.).
The geometry of glass capillary tube in water [Fig.~\ref{f-a}(b)]
was used to simulate the experiments with biomedical relevance
where the photoacoustic signal generation is separated from
detection. \red{The first peak of the time-dependent photoacoustic
signal which corresponds to the fundamental ultrasound signal was
used as a measure of photoacoustic response from the fs
laser-irradiated gold nanosphere colloidal suspensions. Then, the
average of the measured photoacoustic signal intensities was
determined for three experimental trials.}

\subsection{\red{Pre-pulse scattering intensity measurements}}

\red{Near-IR fs-laser pulses (horizontally-polarized) was focused
onto a water-filled quartz cuvette, creating a super-continuum
white light (SWL)  with a broad band emission wavelength of
$\lambda = 300-1000$~nm. The SWL was used as a strobe light to
perform dark field imaging and scattering measurements on
pre-pulse (vertically-polarized) excited gold nanosphere colloidal
suspensions. The pre-pulse energy was maintained at 100~$\mu$J and
its scattering intensity measurements were investigated throughout
the time delay range of 0 to 15~ns. A photodiode was used as a
detector for light scattering intensity measurements. The maximum
scattering light intensity was observed at the wavelength
$\lambda=600$~nm, which was used as a basis for scattering
intensity measurements.}

\section{Results}

\subsection{Photoacoustic intensity enhancement under double-pulse excitation}

\begin{figure}[tb]
\begin{center}
\includegraphics[width=11cm]{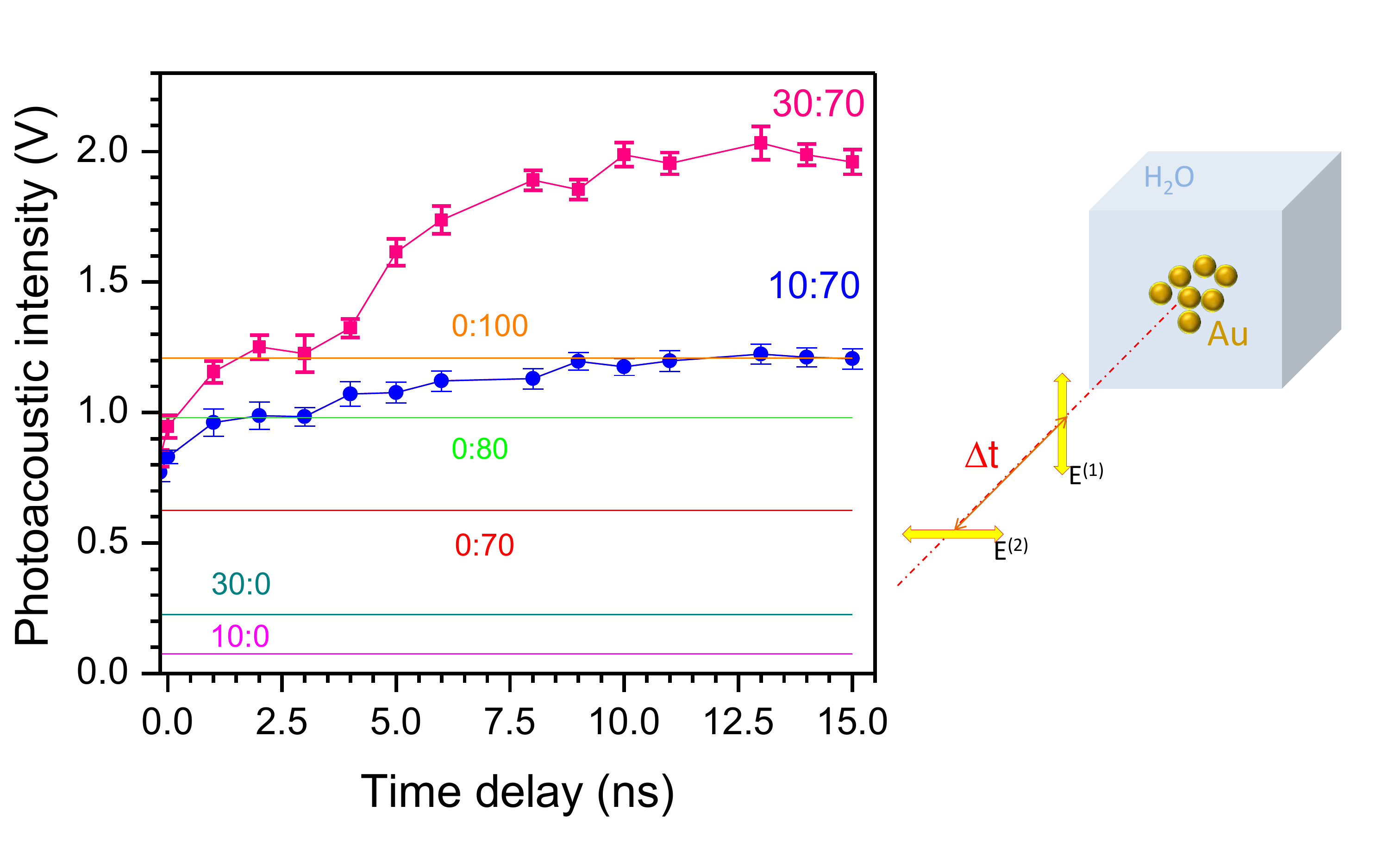}
\caption{Photoacoustic intensity as a function of time delay
between fs-laser pulses vertically (p-pol.) and horizontally
(s-pol.) from $\Delta t = 0$ to 15~ns delay. Pulse energy ratio of
pre-pulse to the main pulse $E_p^{(1)} : E_s^{(2)}$ in $\mu$J is
shown at their approximate maximum signal levels.} \label{f-b}
\end{center}
\end{figure}

The photoacoustic intensity as a function of time delay between
fs-laser pulses vertically (p-pol) and horizontally (s-pol) from
$\Delta t = 0$ to 15~ns is shown in Fig.~\ref{f-b}. The
measurements of photoacoustic intensity were taken under single
pulse (s-pol or p-pol) and double-pulse (s-pol and p-pol)
excitation to Au nanosphere colloidal suspension. A significant
increase in photoacoustic intensity was observed under polarized
double-pulsed excitation ($E_p^{(1)}$: $E_s^{(2)}$= 30:70 $\mu$J
and $E_p^{(1)}$: $E_s^{(2)}$= 10:70 $\mu$J) as the time delay
between pre-pulse $E_p^{(1)}$ and the main pulse $E_s^{(2)}$
increased from 0 to 15 ns. At $E_p^{(1)}$: $E_s^{(2)}$=
30:70~$\mu$J (double-pulse), the photoacoustic intensity reached
up to 3.2, 2.4 and 1.8 times higher than that of $E_s^{(2)}$= 70
$\mu$J (main pulse), $E_p^{(1)}$+$E_s^{(2)}$= 30 +70 $\mu$J (sum
of pre-pulse and main pulse), and  $E_s^{(2)}$= 100~$\mu$J (main
pulse), respectively. When the pre-pulse intensity was decreased
from 30 to 10~$\mu$J, the photoacoustic intensity decayed 1.8
times. The significant enhancement in photoacoustic intensity
under double-pulse excitation is attributed to the efficient
photon energy \red{coupling to nanoparticles} at the nanosecond
time scale due to the initial state of thermal expansion and
generation \red{of nano-bubbles (Sec.~\ref{disco})}.

%__________________________________Fig. 2

%__________________________________Fig. 3
\begin{figure}[t!]
\begin{center}
\includegraphics[width=8cm]{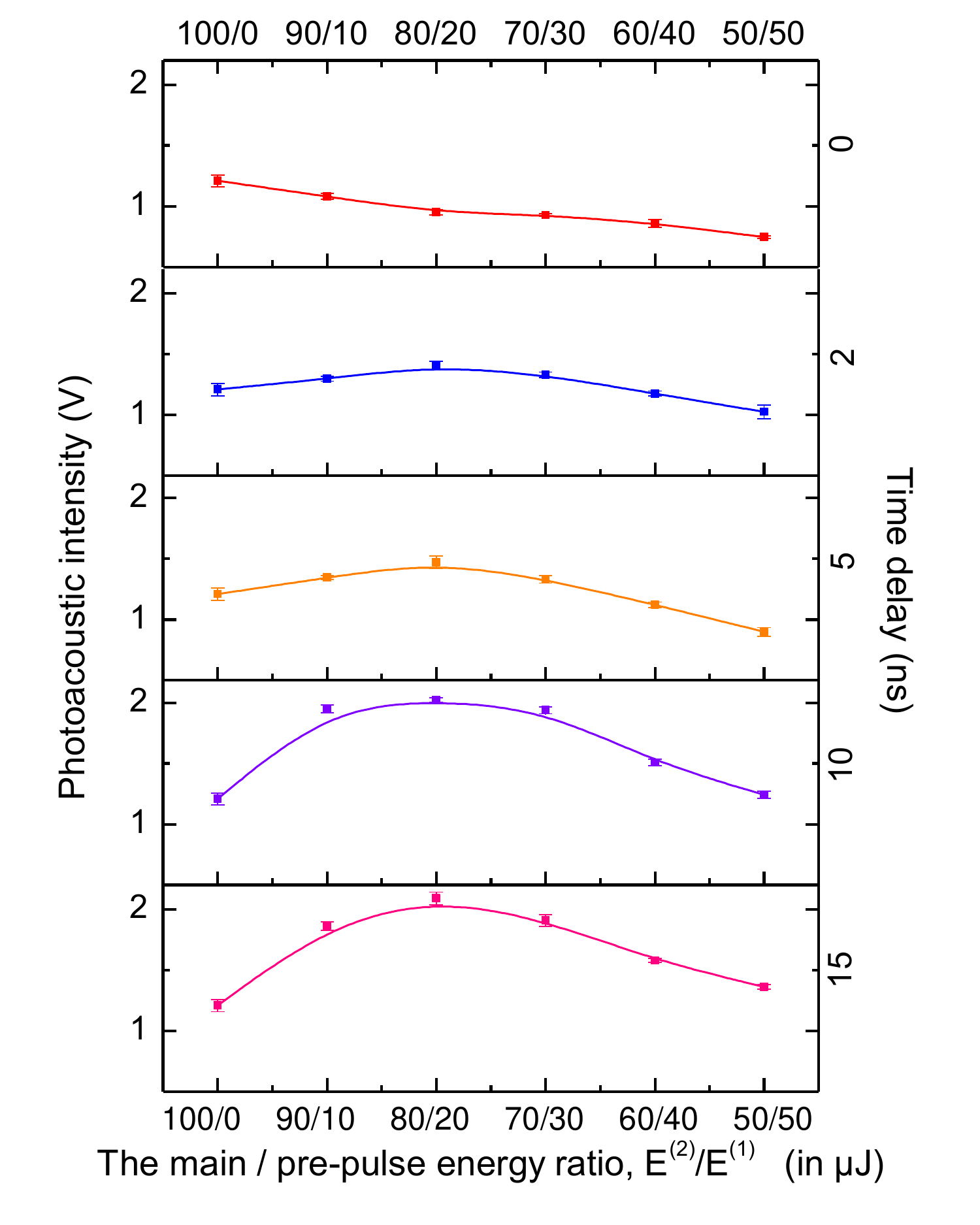}
\caption{Pulse energy ratios with the same total fluence as a
function of photoacoustic intensity at different time delays. The
delay time between the pre-pulse and main pulse was varied from 0,
2, 5, 10 and 15 ns. A total fluence of 100 $\mu$J was used in the
experiments. Lines are drawn as eye guides.} \label{f-c}
\end{center}
\end{figure}

In Fig.~\ref{f-c}, pulse energy ratios for the same total fluence
as a function of photoacoustic intensity at different time delays
($\Delta t = 0, 2, 5, 10, 15$~ns) were systematically
investigated. Different pulse energy ratios of $E_s^{(2)}$:
$E_p^{(1)}$ = 100:0, 90:10, 80:20, 70:30, 60:40 and 50:50 $\mu$J
(same total fluence of 100 $\mu$J) were used to determine the
optimal ratio that could yield to the highest photoacoustic
enhancement. At $\Delta t=0$, the photoacoustic intensity was
linearly dependent on the main pulse energy $E_s^{(2)}$ upon
varying the $E_s^{(2)}$: $E_p^{(1)}$ ratio; highest photoacoustic
intensity at $E_s^{(2)}$: $E_p^{(1)}$ = 100:0 $\mu$J and lowest at
$E_s^{(2)}$: $E_p^{(1)}$ = 50:50~$\mu$J. \red{Photoacoustic
intensity is  dependent on the main pulse energy.} When the time
delay between the main pulse $E_s^{(2)}$ and pre-pulse $E_p^{(1)}$
increase from $\Delta t$ = 2 to 15~ns, a noticeable peak at
$E_s^{(2)}$: $E_p^{(1)}$ = 80:20 $\mu$J starts to grow from
$\Delta t$ = 2~ns and reaches its maximum intensity at $\Delta t$
= 15~ns.  The $E_s^{(2)}$: $E_p^{(1)}$ = 80:20 $\mu$J was
\red{found} to be the optimal ratio with the highest enhancement
in photoacoustic intensity. \red{Accordingly, with $E_s^{(2)}$:
$E_p^{(1)}$ = 80:20~$\mu$J at $\Delta t$ = 15~ns, the maximum
enhancement in photoacoustic intensity was achieved which is
ascribed to the  bubble generation in the nanosecond time scale.}
The photoacoustic signal growing in two recognizable shorter
$\sim$~2 ns and longer $\sim$~15~ns stages
\red{[Fig.~(\ref{f-b})]}, which is consistent with light
scattering data (Sec.~\ref{scatt}). \red{Light scattering i}s
sensitive to the volume of the optically excited \red{region,
i.e., nano-bubbles and nanoparticles. The saturation of
photoacoustic signal reached at the end of 15~ns was at the limit
of the utilised delay line, however, further increase is not
expected due to the observed scattering decay with the $\sim
17$~ns time constant (Sec.~\ref{scatt}) and the known strong
damping of nano-bubble oscillation and their short lifetime of up
to few nanoseconds when nanoparticles of similar size were
used~\cite{Siems2011}}.

%__________________________________Fig. 4
\begin{figure}[tb]
\begin{center}
\includegraphics[width=10cm]{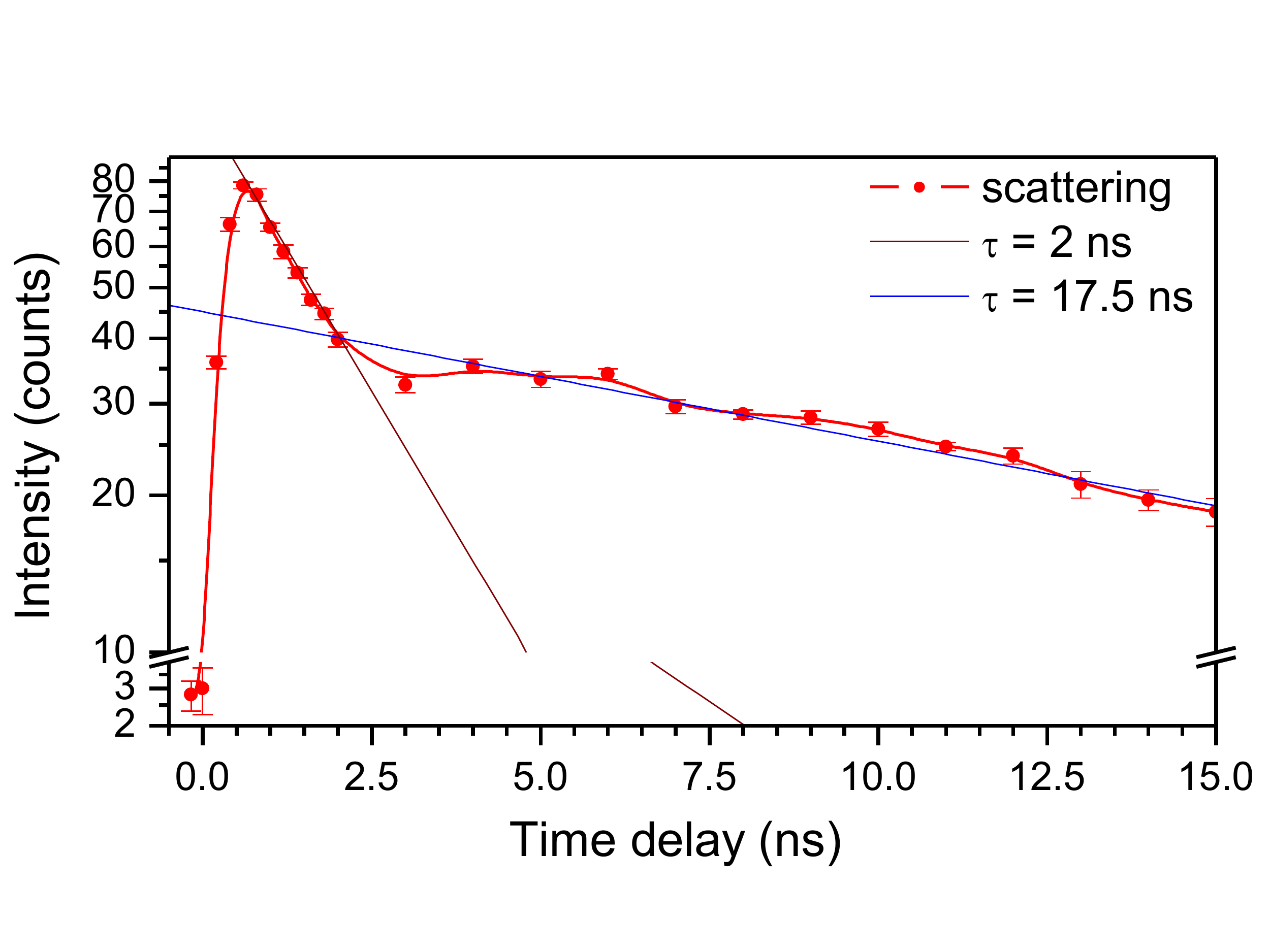}
\caption{Pre-pulse scattering intensity at the $\lambda$ = 600~nm
wavelength measured from 0 to 15~ns time delay between the pulses.
A supercontinuum white light generated by fs-laser was used as
strobe light to perform dark-field imaging and scattering
measurements. Two single exponential decays with time constants
$\tau = 2, 17.5$~ns are shown as best fits.} \label{f-d}
\end{center}
\end{figure}

\subsection{\red{Bubble generation by f}emtosecond
double-pulse}\label{scatt}

In the early stage of excitation by fs-laser pulses which occurs
in several picosecond time scale after electron-ion
thermalization, pre-pulse plays an important role in the thermal
excitation dynamics. The pre-pulse of the double delayed fs-laser
pulses \red{clearly} affected the total laser energy coupling into
gold nanoparticles and finally the ablation characteristics. To
study the behavior of pre-pulse from 0 to 15~ns, the pulse energy
was \red{kept at 100~$\mu$J} and the dynamics was investigated.
Figure~\ref{f-d} shows the pre-pulse scattering intensity as a
function of time delay between the pulses. At 0 to 2~ns, a rapid
increase in the scattering intensity was observed due to the
\red{nanoparticle} ablation and \red{nano-bubble} generation. A
single-exponential decay with 17.5~ns time constant was observed
\red{ after initial faster 2~ns decay}.

%As first step, a \red{nano-bubble} formation dynamics can be
%estimated from speed of sound in water $v_s = 1.484$~km/s which
%travels $l_s = v_s\times\Delta t \simeq 22.3~\mu$m during $\Delta
%t = 15$~ns time and $3~\mu$m in 2~ns (a photodiode response time
%was $\sim 0.5$~ns).

For the focusing objective lens with numerical aperture $NA =
0.28$, the diameter of the focal spot is $2w_0 = 1.22\lambda/NA =
3.5~\mu$m. The 2~ns slope in the light side-scattering transient
\red{[Fig.~(\ref{f-d})]} can be considered as \red{nano-bubble}
initiation out of the focal volume. Theoretical axial extent of
the focal volume can be estimated as a double Rayleigh length
$2z_R = 2n\frac{\lambda}{NA^2} = 27.1~\mu$m and is close to the
estimate made above for pressure traverse time of 17.5~ns; $n =
1.33$ is the refractive index of water. In the previous single
pulse excitation photoacoustic and X-ray generation experiments,
the side view images of the optically excited expanding volume had
an axial extent of $\sim 30~\mu$m at $E^{(1)} =
30~\mu$J~\cite{Masim2}.

\section{\red{Discussion}}\label{disco}

\red{Separation between gold nanoparticles in solution was only
$\sim 1.4~\mu$m which is smaller than $30~\mu$m when growth of
nano-bubbles is independent~\cite{Nakajima2016}. Size of
nano-bubbles measured with X-ray scattering and shadowgraphy was
around 1~$\mu$m diameter for $\sim$40-nm-diameter nanoparticles at
typical range of pulse fluences 0.1-0.3~J/cm$^2$ when nanosecond
laser pulses were used~\cite{Siems2011,Boutopoulos2015}. For
femtosecond laser pulses, the threshold of  bubble formation is
only twice lower for  the optimum size for the lowest threshold of
nano-bubble initiation with a wide minimum at 40-60~nm
diameters~\cite{Metwally2015,Siems2011}. It is defined by the
plasmonic scattering and absorption contributions to extinction
and Kapitza resistance at the nanoparticle-water interface which
is responsible for a significant overheating of the nanoparticle.
The observed 2~ns time constant in light scattering is consistent
with the initial stages of nano-bubble growth reported in
literature~\cite{Boulais2013,Lachaine2014,Plech2004}. The long
$\sim 17$~ns decay is caused by bubble growth which has typical
times of 15-25~ns and is longer for the higher
fluence~\cite{Boutopoulos2015}. At the used high density solution,
the  pressure waves from adjacent bubbles (nanoparticles) inhibits
bubble growth~\cite{Nakajima2016}, but provides a homogenised
volume of high pressure as a source of photoacoustic signal.}

%__________________________________Fig. 5
\begin{figure}[tb]
\begin{center}
\includegraphics[width=14.cm]{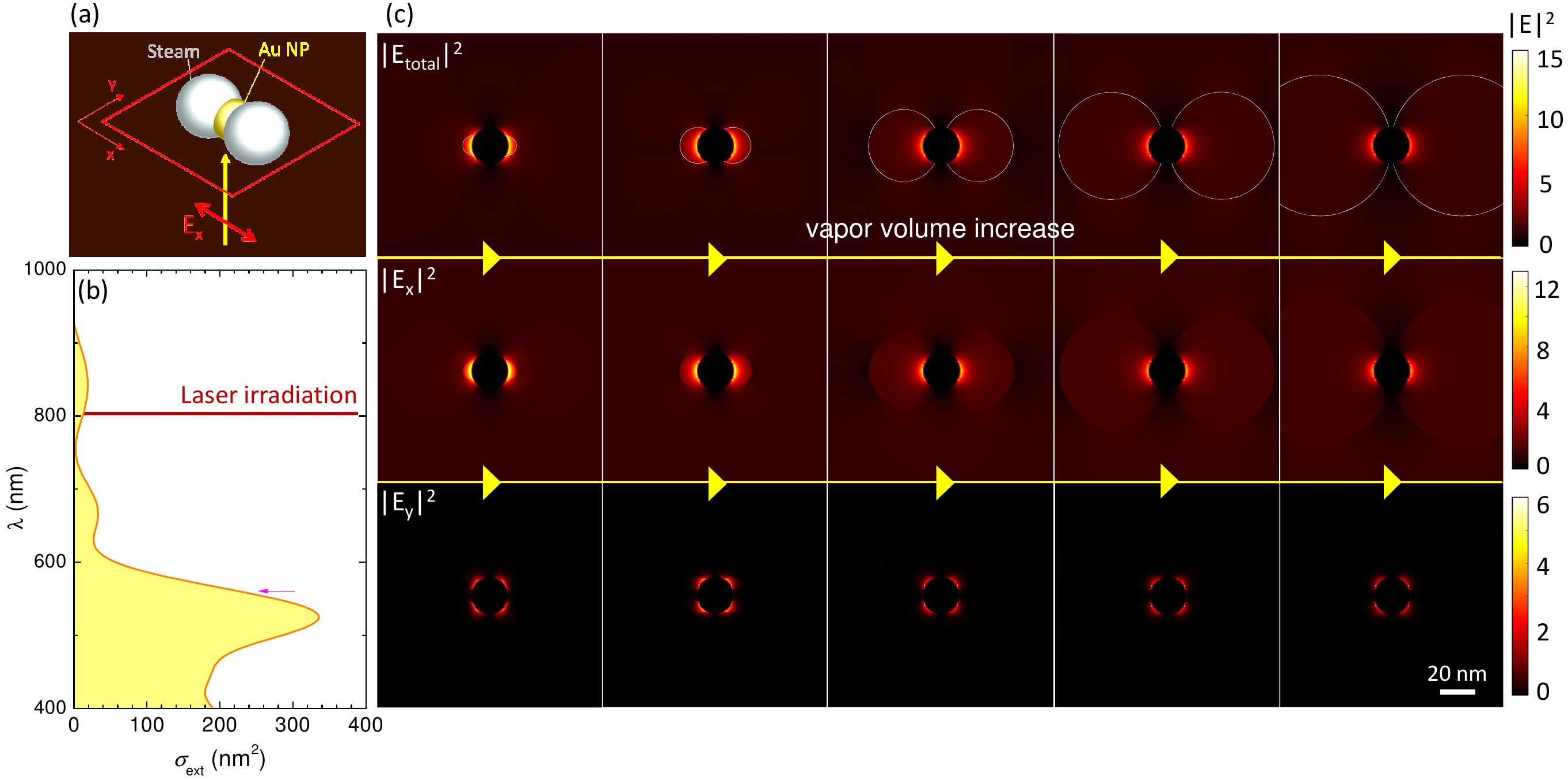}
\caption{(a) Schematic representation of the simulation geometry,
with 800~nm wavelength light incident along z-axis and is linearly
polarized along x-axis. (b) Simulated extinction cross section,
$\sigma_{ext}$, spectrum of the 20~nm diameter Au nanoparticle.
The arrow marks the wavelength where two-photon absorption has
maximum $\sim 0.7\lambda_{ex} = 560$~nm~\cite{10oe10209}. (c)
Evolution of the absolute ($|E_{total}|^2$) as well as the
$|E_x|^2$ and $|E_y|^2$ component electric field intensity
profiles in the x-y plane around the nanoparticle as the bubbles
expand. The intensity of the $|E_z|^2$ components is three orders
of magnitude lower, hence their plots are omitted.} \label{f-fdtd}
\end{center}
\end{figure}

\red{Another feature specific to this study is very high pulse
fluence $\sim 1$~kJ/cm$^2$ far exceeding that typical for
nano-bubble formation and cell perforation at $<
0.1$~J/cm$^2$~\cite{Lachaine2016}. At such high intensity, air
breakdown at the air-water interface, white light continuum
generation, and filamentation are all contributing to significant
reduction of the light intensity reaching the nanoparticle. It was
established that once 0.24~J/cm$^2$ fluence is exceeded, a
repeated irradiation of the same nanoparticle did not produce
nano-bubbles~\cite{Boutopoulos2015}. Particle reshaping and
resizing was observed. We used flow of nanoparticles and the
entire volume of the irradiated solution was smaller than 10\%
after entire experiment. Extinction spectra measured before and
after photoacoustic measurements had the same spectral shape.
Ablation and disintegration of nanoparticles can explain the
observation. Initiation of nano-bubbles by plasma and electronic
emission at the surface of nanoparticle was demonstrated for
fs-laser pulses as an alternative to thermally initiated spinoidal
water decomposition at strong overheating conditions typical for
nanosecond pulsed irradiation~\cite{Lachaine2014}. Surface plasma
emission from the light field enhancement locations (hot-spots) is
relevant mechanism  at the used high irradiance by fs-laser pulses
employed in our study. Creation of hot-spots with light field
enhancement by several times was modeled numerically.}

Figure~\ref{f-fdtd} shows finite-difference time-domain (FDTD)
simulation results of light intensity distribution for a 20~nm
diameter Au nanoparticle suspended in water, assuming that
plasmonic hot-spots induce \red{nano-bubble} formation around
localized high intensity regions, hence, the dipolar pattern of
the vapor volume. Initial stages of \red{nano-bubble} generation
are taken and the light field enhancement is shown for the major
components of the E-field. The time \red{estimate} for pressure
wave travel 20~nm in water takes $\sim 13.5$~ps \red{at velocity
of sound and can be few times faster at shock wave conditions
typical for such experiments~\cite{Boulais2013,Siems2011}}; this
can be considered as an bubble initiation time. It is evident
\red{[Fig.~(\ref{f-fdtd})]} that, there is no new energy
deposition possibilities due to opening of vapor volumes around
nanoparticles via an augmented light enhancement nor due to a
resonant absorption at the interfaces liquid-vapor and gold-vapor
which can be important in polarized double-pulsed
experiment~\cite{08oe12650}; calculations were also carried out
for spherical and toroidal volumes, however, there were no
significant differences. \red{However, this modeling is not
capturing presence of plasma and conditions of white light
continuum. In actual experiments additional energy deposition
channels also exist via } two photon absorption
 \red{TPA} [Fig.~\ref{f-fdtd}(b)] \red{and} opens an efficient energy
deposition. \red{T}he extinction cross section has a major
component \red{at TPA wavelength [Fig.~\ref{f-fdtd}(b)]. Due to
absorption  dominance in extinction }$\sigma_{ext} = \sigma_{abs}
+ \sigma_{sc} \simeq \sigma_{abs}$ \red{the} scattering and
reflection for particles with diameter smaller than $\sim$40~nm
\red{are considerably weaker}~\cite{Masim1}.

\red{Future studies of high-density solutions of nanoparticles
within a small focal volume of excitation should provide optimized
solutions for photoacoustic sources.}

\section{Conclusion}

Double-pulsed fs laser irradiation to gold nanosphere colloidal
suspension demonstrated a significant increase in photoacoustic
intensity from 0 to 15~ns time scale. An efficient enhancement
factor of $\sim 2$ was achieved when the pre-pulse energy is
20-30\% of the total energy. The initial stage of thermal
expansion and \red{bubble} generation in the nanosecond time scale
were maximized under double-pulsed femtosecond
excitation, leading to a significant photoacoustic enhancement.
The pre-pulse of the time-delayed fs-laser pulses affected the total
laser energy coupling to gold nanoparticles and thereby enhancing
the ablation and photoacoustic characteristics.

\section*{\red{Acknowledgment}}

TY acknowledges the partial support of Murata Foundation.

\section*{\red{Funding}}

SJ is grateful for partial support via the Australian Research
Council DP170100131 Discovery project and by the nanotechnology
ambassador fellowship program at the Melbourne Centre for
Nanofabrication (MCN) in the Victorian Node of the Australian
National Fabrication Facility (ANFF).

\end{document}